\documentclass{article}

\usepackage[english]{babel}
\usepackage[
    natbib=true,
    style=numeric,
    sorting=none
]{biblatex}

\usepackage{arxiv}

\usepackage[utf8]{inputenc} 
\usepackage[T1]{fontenc}    
\usepackage{amsfonts}       
\usepackage{graphicx}
\usepackage{caption}
\usepackage{subfig}

\title{A stochastic differential equation based algorithm to simulate laser speckles for deep tissue blood flow imaging applications}

\author{
Murali K \\
 Department of Biosciences and Bioengineering,\\ Indian Institute of Technology Bombay, India \\
   \And
 Hari M Varma \\
 Department of Biosciences and Bioengineering,\\ Indian Institute of Technology Bombay, India \\
  \texttt{harivarma@iitb.ac.in} \\
}

\begin{document}
\maketitle
\begin{abstract}
We present  an intensity speckle simulation algorithm based on stochastic differential equations. Intensity speckles are generated with a negative exponential distribution and an exponential auto-correlation decay. The mean of the distribution is spatially varying dictated by photon diffusion to take into account of diffuse speckles. The algorithm is validated using simulation studies for both surface and deep tissue blood flow with potential applications in diffuse correlation spectroscopy.
\end{abstract}


\section{Introduction}

Laser speckles have been used for quantifying both surface and deep tissue blood flow by appropriate laser illumination on tissue and measuring the scattered intensity \cite{boas2010laser, durduran2010diffuse}. The scattered intensity is statistically quantified using either auto-correlation or speckle contrast, which is then related to blood flow using appropriate models. In addition to conventional laser speckle contrast imaging (LSCI) \cite{briers1982retinal} and diffuse correlation spectroscopy (DCS) \cite{durduran2010diffuse}, several variants of laser speckle imaging system have been reported in the recent times namely multi-exposure speckle imaging (MESI) \cite{parthasarathy2008robust}, diffuse speckle contrast analysis \cite{bi2013deep}, speckle contrast optical spectroscopy (SCOS)\cite{valdes2014speckle}, Multi speckle DCS (M-DCS)\cite{murali2020multi} and Interferometric DCS (I-DCS) \cite{zhou2018highly}). To this end, a means of simulating speckles for the above applications is necessary.

Several methods that were reported in the past to simulte laser speckles for surface blood flow (as in LSCI) are given in Ref \cite{kirkpatrick2008detrimental,duncan2008copula,song2016simulation,rabal2018dynamic}. One of the common approches is to use fast Fourier transform of the phase matrix \cite{kirkpatrick2008detrimental}, which can only generate independent speckle pattern, without any correlation statistics. In Ref \cite{duncan2008copula} , Duncan et al., proposes a method based on Copula algorithm for generating speckle sequences for a given auto-correlation model that utilizes the concept of the quantile function and direct Fourier transform. However this method was not utilzed for generating spatially varying speckle patterns. The concept of coherent imaging principles is used to generate spatially varying speckle images using predefined correlation coefficient matrices in Ref \cite{song2016simulation}.  A comprehensive review of some of the methods used for generating speckles in general can be found in chapter 3 of Ref \cite{rabal2018dynamic}. 

In this paper, we present an algoirthm to simulate intensity speckles with a known probablity density function (pdf) and auto-correlation by solving a stochastic differential equation (SDE).  A SDE, in simpler terms, is an ordinary differential equation with a stochastic term, whose solution is a stochastic process. The SDE have several applications in various fields that include but not limited to, chemistry, finance, epidemiology, mechanics and microelectronics \cite{braumann2019introduction}\cite{yin2015improved}. Here for the first time, we apply the concept of SDE to generate laser speckles for blood flow imaging application. Additionally, we also generate spatially varying diffuse speckles by incorporating the solution of photon diffusion equation along with SDE with a potential of application in deep tissue blood flow imaging. 

\section{Theory}

Consider a SDE which consist of an ordinary differential equation along with a stochastic term given as,
\begin{equation}
dx(t)=a(x(t),t)dt + b(x(t),t)dW(t); \\
x(0)=x_0.
\end{equation}
Here $a$ and $b$ are called the drift and diffusion terms respectively and $W(t)$ is the Weiner process (or Brownian motion) such that $dW(t) \sim \sqrt{\delta t} N(0,1)$. It is well known that functions a and b satisfies the Fokker-planck equation \cite{braumann2019introduction}, which is given as, 
\begin{equation}
\frac{\partial p}{\partial t} = -\frac{\partial }{\partial x} (a p) + \frac{\partial^2}{\partial x^2} (\frac{b^2}{2} p).
\end{equation}

Here $p$ is the probability density function (pdf) of random variable $x(t)$. We assume that the random process is stationary, so that a and b implicitly depends on time while p is independent of time, which resulting in,

\begin{equation}
\frac{\partial(ap)}{\partial x} = \frac{1}{2} \frac{\partial^2(b^2 p)}{\partial x^2}.
\end{equation}
For x(t) to represent intensity speckles, it should have a negative exponential pdf, i.e, p(x)= $\frac{1}{\mu}exp(-\frac{x}{\mu})$, where the mean and the standard deviation are equal to $\mu$. Additionally, the auto-correlation of the intensity speckles obeys exponential decay \cite{goodman2007speckle}. The usual procedure to enforce this in the solution of equation (1) is to appropriately choose the parameters 'a' and 'b' using equation (3). It can be seen that, the terms $a(x(t))=-\alpha(x-\mu)$ will ensure a correlation decay of $
\mathcal{T}(\tau)=e^{-\alpha\tau}$, which results in the equation $dx(t)=-\alpha(x-\mu)dt + b(x(t))dW(t)$ \cite{ gardinerstochastic,zarate2016construction}. This predetermined 'a' is now substituted in equation (3) along with the exponential pdf to deduce 'b' as follows: $b^2(x(t))=   \frac{2}{p(x)}\int_0^x a(x(s)) p(s) ds          =   \frac{2}{p(x)}\int_0^x -\alpha(x(s))-\mu) p(s) ds          =   2\alpha\mu x.$

The resulting SDE (with a change of notation $I \equiv x$), is given by 
\begin{equation}
dI(t)=-\alpha	(I(t)-\mu) dt + \sqrt{2\alpha\mu I(t)} dW(t),
\end{equation}
whose solution gives the intensity speckles with aprior pdf and auto-correlation. Note that the above SDE as in equation (4) is known as Cox-Ingersoll-Ross (CIR) model ($dX(t)=-\alpha(X-\mu)dt + \sigma \sqrt{X(t)} dW$) in mathematical finance, where X(t) represent the interest rate. Clearly for our case, I(t) has to be positive, which is ensured by Feller's condition, given as $\frac{2\alpha\mu}{\sigma^2} \geq 1$. In our case, $\frac{2\alpha\mu}{\sigma^2}=\frac{2\alpha\mu}{2\alpha\mu}=1$ and hence $I(t) > 0$ for all t.  The resulting SDE in intensity is solved using Milstein scheme \cite{zarate2016construction, higham2001algorithmic, gardinerstochastic}, given by,
\begin{equation}
I_{n+1} = I_n + a(I_n) \Delta t + b(I_n)\Delta W_n + 0.5 b(I_n)b^{'}(I_n)[(\Delta W_n)^2 - \Delta t], 
\end{equation}
with $I_0=\mu$ and where n is the index of time step, 

Depending on the application, we can take autocorrelation to be independent on spatial co-ordinates 'r' (in an homogeneous sample with uniform illumination as in LSCI/ MESI) or dependent on 'r' as in diffuse speckles (DCS, DSCA, SCOS, M-DCS and I-DCS).  Thus for LSCI, we have $g_L(\tau)=e^{-\alpha \tau}$, $\alpha=1/\tau_c$, where $\tau_c$ is the characteristic decay time. The normalized intensity auto-correlation obtained by the SDE is denoted as $\tilde{g_2}$, defined as $\tilde{g_2} \equiv g_2(\tau)-1 = e^{-2\alpha\tau}$, which can be deduced from the Siegert's relation \cite{boas2010laser}. For simplicity, we define $\tilde{\alpha}=2\times\alpha$, hence the auto-correlation $\tilde{g_2}^L(\tau)=e^{-\tilde{\alpha} \tau}$. 

For DCS, we use an infinite medium solution of CDE which is of the form $g_{1}^{D}(\tau) = \frac{e^{-K_{1}r}}{r}/ \frac{e^{-K_{2}r}}{r}$;  where $K_{1}=\sqrt{3\mu_{a}\mu_{s}^{'} + 6 \mu_{s}^{'} k_{0}^{2} D_{B} \tau}$ and $K_{2}=\sqrt{3\mu_{a}\mu_{s}^{'}}$; Here, $\mu_{a}$,$\mu_{s}^{'}$ are the absorption and reduced scattering coefficients respectively, $k_{0}$ is the wave number and $D_{B}$ is the particle diffusion co-efficient.
Since, we have assumed a simple auto-correlation decay of the form $e^{-\alpha\tau}$ for $I(t)$, we use a binomial approximation to simplify the normalized field auto-correlation of DCS to get $\tilde{g_1}^D=exp(-\alpha_{D}\tau)$, where $\alpha_D= \frac{6\mu_{s}^{'2} k_{0}^{2} D_{B} r}{2\sqrt{3 \mu_{a} \mu_{s}^{'}}}$. Hence the resulting normalized intensity auto-correlation from SDE is given as $\tilde{g_2}^D(\tau)=exp(-\tilde{\alpha_D}\tau)$, where $\tilde{\alpha_D}=2\times \alpha_D$.

\section{Simulation results and discussion}

We have solved equation (4) for intensity speckles I(t) by implementing Milstein's scheme given in equation (5) in MATLAB. To validate the method, we have also generated the normalized auto-correlation function $\tilde{g_2}$ obtained using I(t) generated by the above SDE for different $\alpha$ values. The minimum time denoted as $t_{min}$ in equation (4) was fixed as $10^{-6} \times 1/\tilde{\alpha}$, so that the entire decorrelation curve can be captured based on error analysis given in Ref \cite{murali2019recovery}. The maximum t denoted as $t_{max}$ was fixed as $\gamma \times 1/\tilde{\alpha}$, wherein by error analysis $E(\tilde{\alpha}) \equiv$ mean(actual $\tilde{\alpha}$ - fitted $\tilde{\alpha}$) of 50 trials as given in Fig 1(a), we have estimated that $\gamma=3$ gives optimal result. The $\mu$ value was taken as $2 \times 10^{-3}$ and the initial value $I_0$ was fixed as $\mu$. The $randn$ function in MATLAB was used to generate the random numbers between 0 and 1 to obtain $dW(t)$.

The validity of $\tilde{g_1}^D(\tau)$ is dictated by the binomial approximation. The original $g_1^D$ and $\tilde{g_1}^D$ for different r is shown in Fig 1(b). The error, E $\equiv$ (Fitted $D_B$-Actual $D_B$)/(Actual $D_B$) $\times 100$, decreases as r increases as seen from the inset of Fig 1(b). The error was found to be around $21\%$ in terms of flow for r=1 cm and around $12\%$ for r=2 cm.

\begin{figure}
\subfloat[]{\includegraphics[width = 3in]{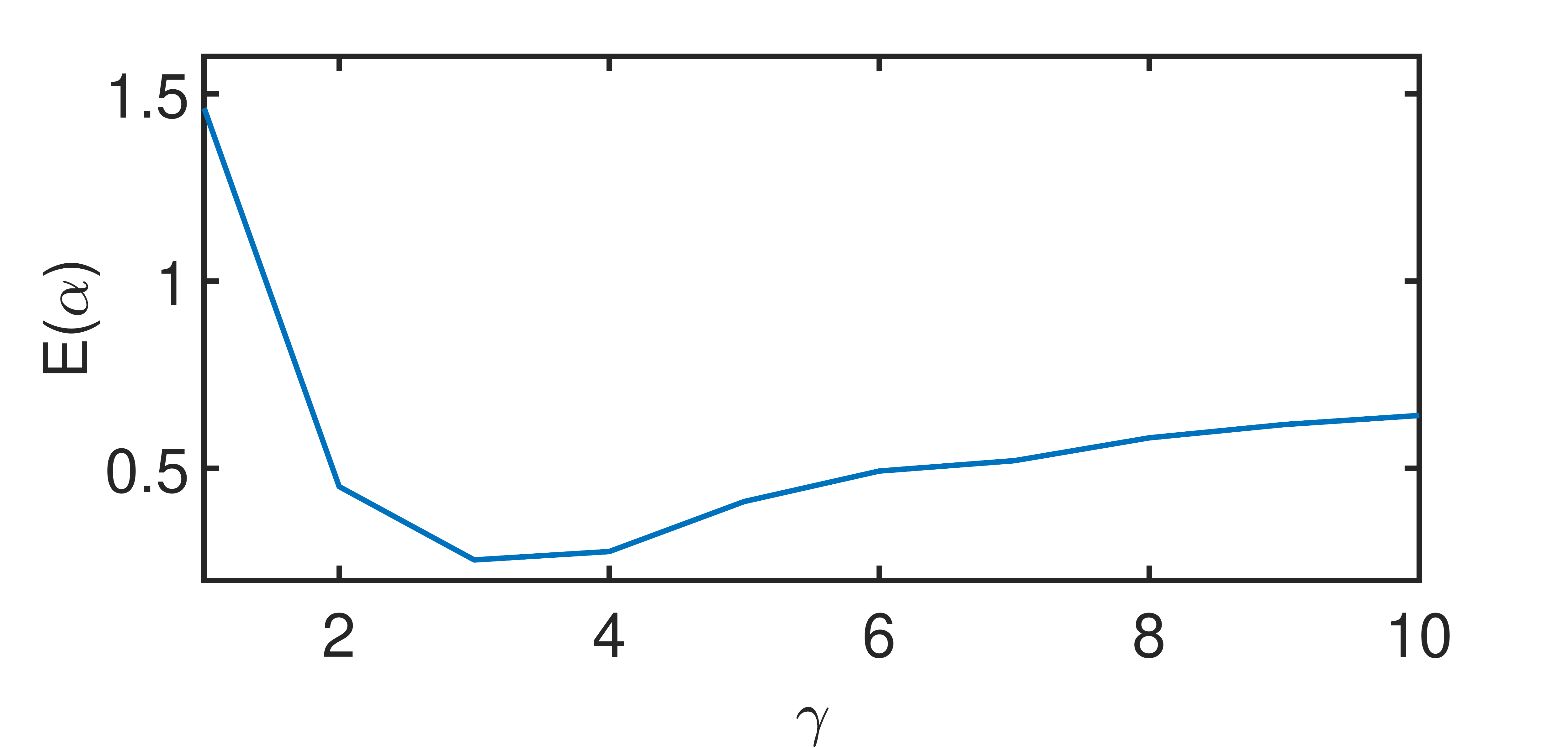}}
\subfloat[]{\includegraphics[width = 3in]{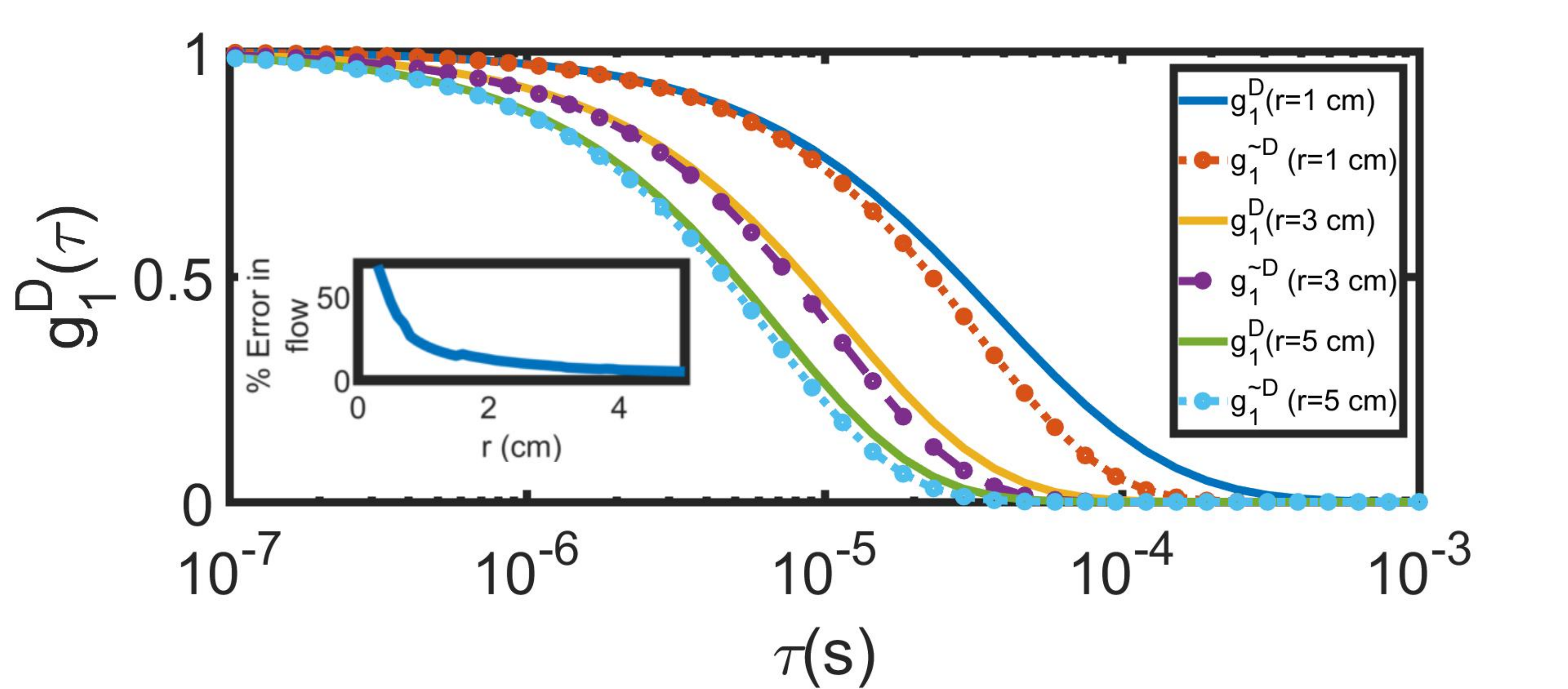}} \\
\caption{Figure (a) shows the error in $\alpha$ obtained when using different values of $\gamma$ and from the plot the $\gamma$ was fixed as 3, wherein the error is minimal. Figure (b) shows the validity of the Binomial approximation of $g_{\tilde{D}}$ for different source detector separation r. The inset plot shows the error as a function of r. }
\label{numerical}
\end{figure}

For surface blood flow, the auto-correlation is given as $\tilde{g_2}^L = exp(-\tilde{\alpha} \tau)$. The intensity speckles generated are shown in Fig 2(a). The corresponding auto-correlation $\tilde{g_2}^{~L}$ was generated for two $\tilde{\alpha}$ values of 1000 and 10000 is shown in Fig 2(b). The auto-correlation was generated for 10 independent trials and the mean and the variance is plotted in Fig 2(b). It can be seen that the auto-correlation generated by SDE is in reasonable comparison to the desired auto-correlation. The auto-correlation was fitted against the theoretical $g_{2}^{~}$ model and the fitted values are reported in the Fig 2(b). The histogram of the intensity speckles are also plotted in Fig 2(c), which follows the negative exponential pattern.

\begin{figure}
\subfloat[]{\includegraphics[width = 3in]{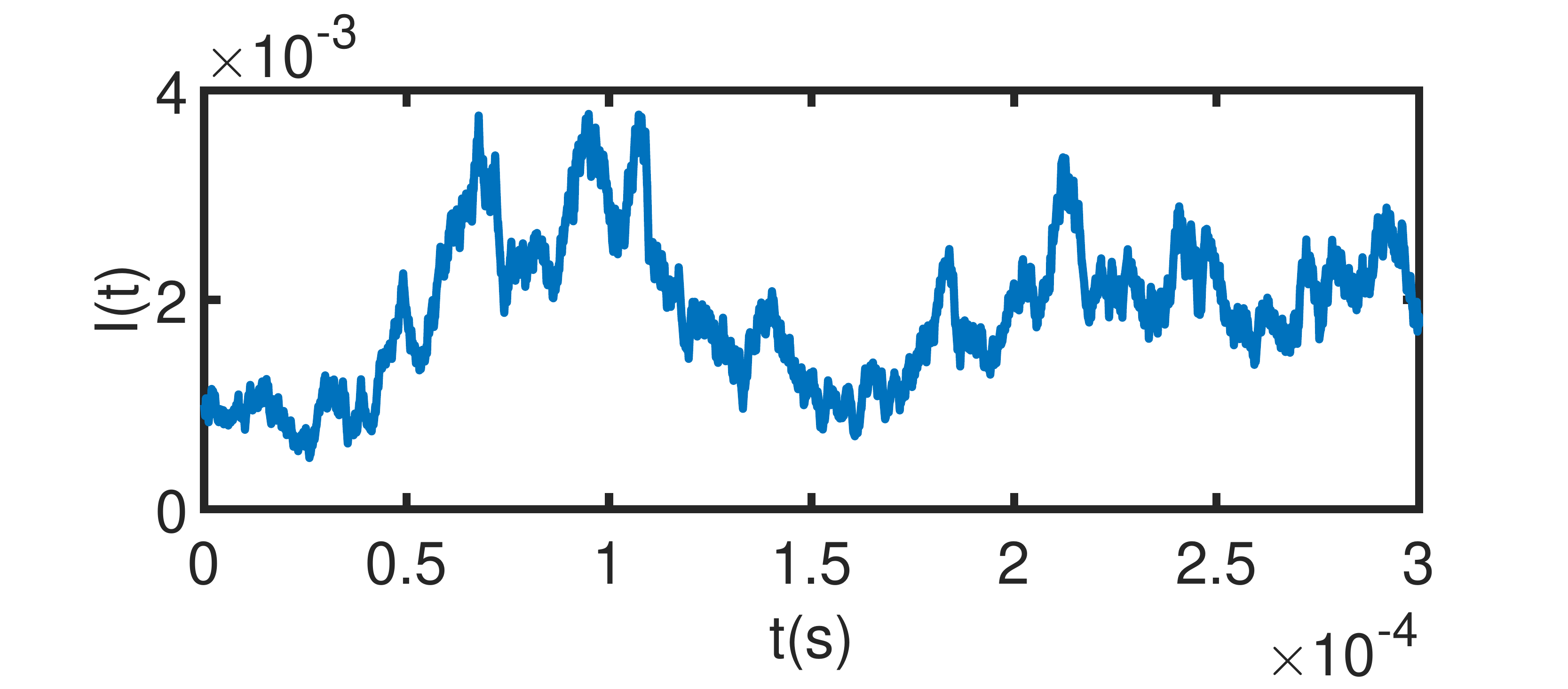}} 
\subfloat[]{\includegraphics[width = 3in]{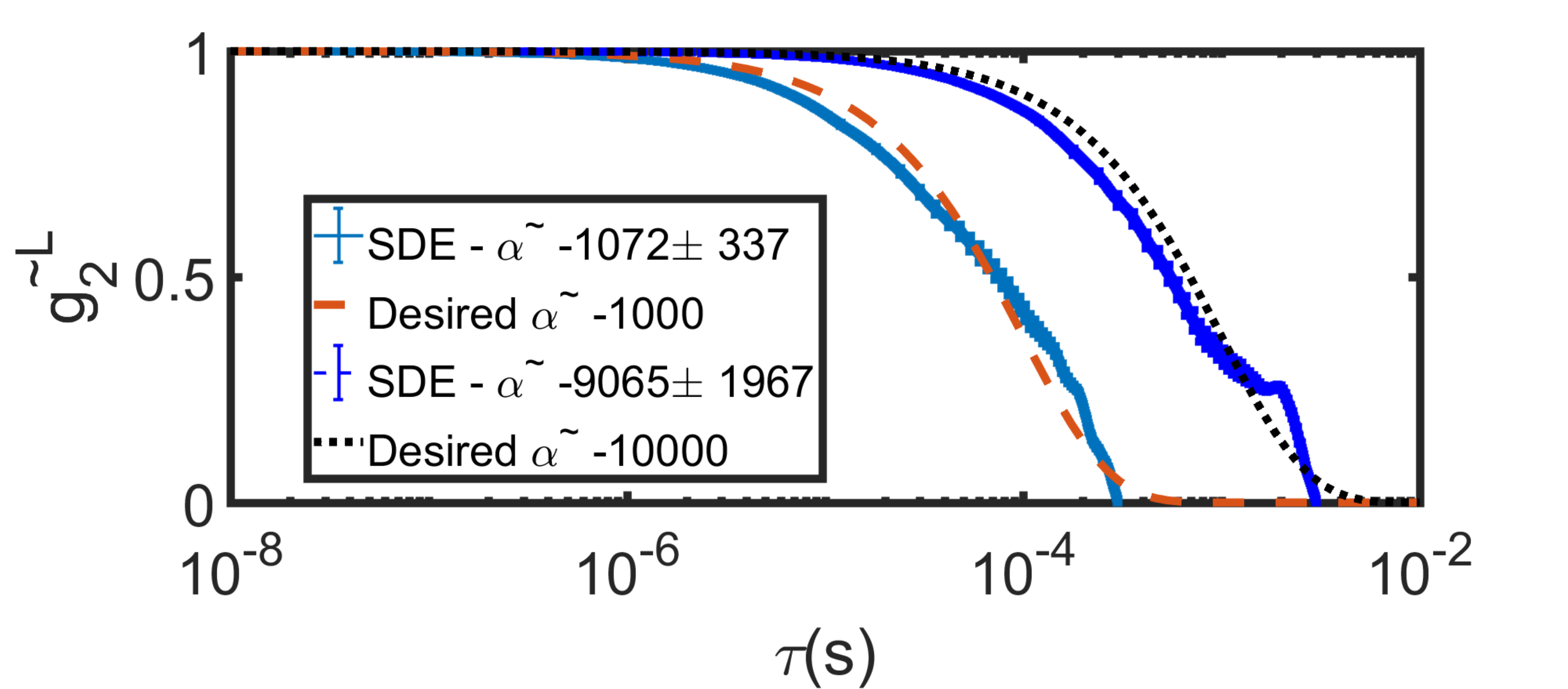}}\\
\subfloat[]{\includegraphics[width = 3in]{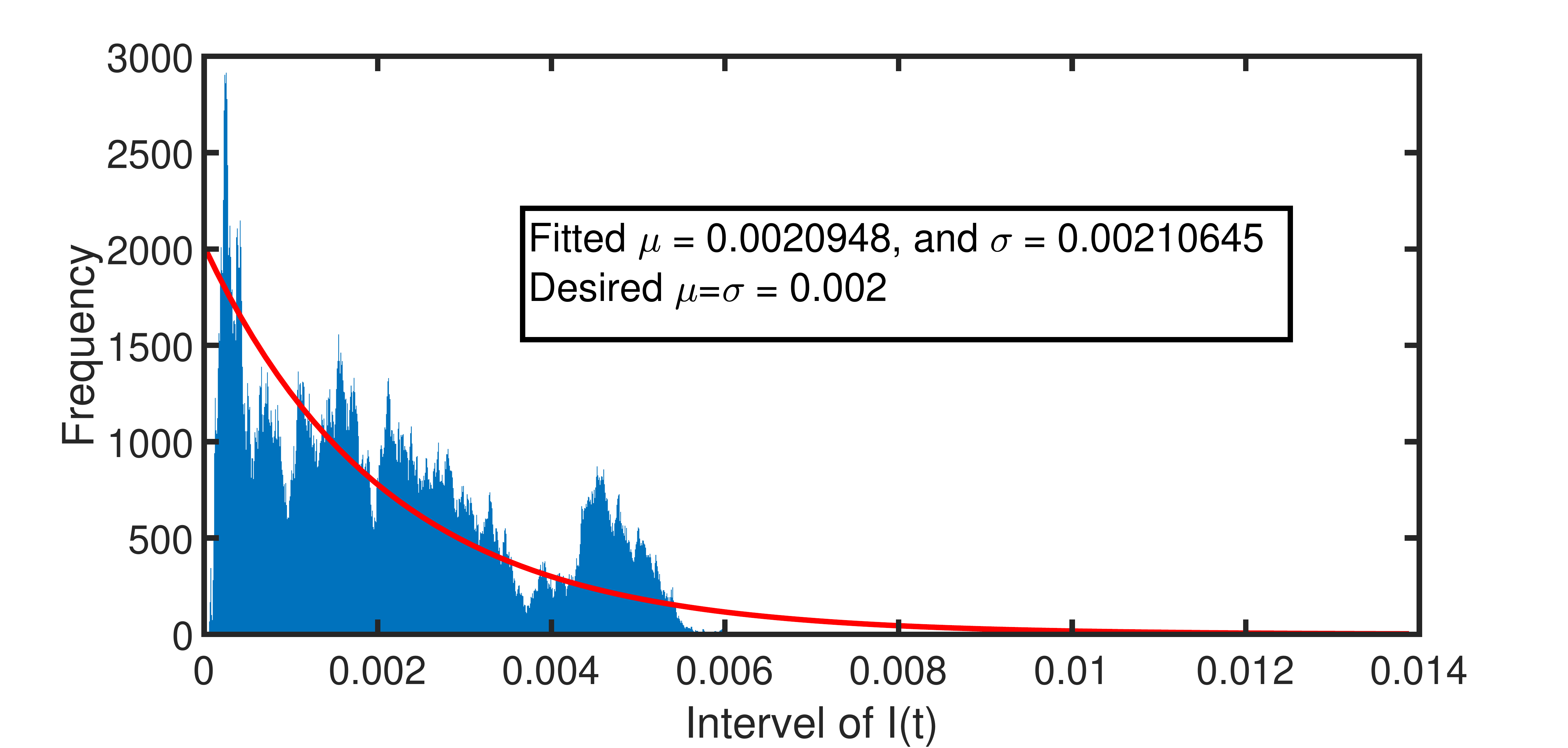}}
\caption{Figure (a) shows the intensity speckles I(t) generated by using SDE. Figure (b) shows the auto-correlation curve $\tilde{g_2}^{L}$ obtained for two different $\tilde{\alpha}$ values and are compared with the desired $\tilde{g_2}^{L}$ for surface blood flow. The mean and the variance of auto-correlation over 10 trials is shown Figure (b). The histogram of the I(t) is shown in Figure (c).}
\label{numerical1}
\end{figure}

For the deep tissue blood flow, $\tilde{\alpha_D}$ was estimated and the corresponding intensity speckles was generated. The $\mu_{a}$,$\mu_{s}^{'}$ and $D_{B}$ used were $0.1cm^{-1} , 12cm^{-1}$ and $1\times10^{-8} cm^2/s$ respectively. The results of $\tilde{g_2}^{D}$ for two different SD separations of $r =1 cm$ and $r=3 cm$ is given in Fig 3(a). It can be seen that they are in reasonable comparison with the desired $g_{2}^{~}$. A $100 X 100$ pixel based image with resolution of $0.03$ cm was created. For each pixel with a given spatial co-ordinate r, the $\mu$ was fixed using the photon diffusion model $\mu(r)=\frac{e^{-\sqrt{-3\mu_a \mu_s^{'}}r}}{r}$ \cite{durduran2010diffuse} and the $\tilde{g_2}^{D}$ was obtained by solving SDE given in equation 5. The log of the intensity speckles ($log(\phi)$) is shown in Fig 3(b) at t=4 $\mu s$ and the corresponding profile plot of $log (\phi.r)$  is shown in Fig 3(c), whose slope is determined by optical properties. The expected slope is $\sqrt{3\mu_{a}\mu_{s}^{'}}$=1.89. The $\tilde{g_2}^D$ obtained using SDE at $\tau = 4 \mu s$ and $\tau = 40 \mu s$ is given in Fig 3(d) and Fig 3(e) respectively. It can be seen from Fig 3(d) and 3(e) that the $\tilde{g_{2}}$ decreases as $\tau$ and $r$ increases.

\begin{figure}
\subfloat[]{\includegraphics[width = 3in]{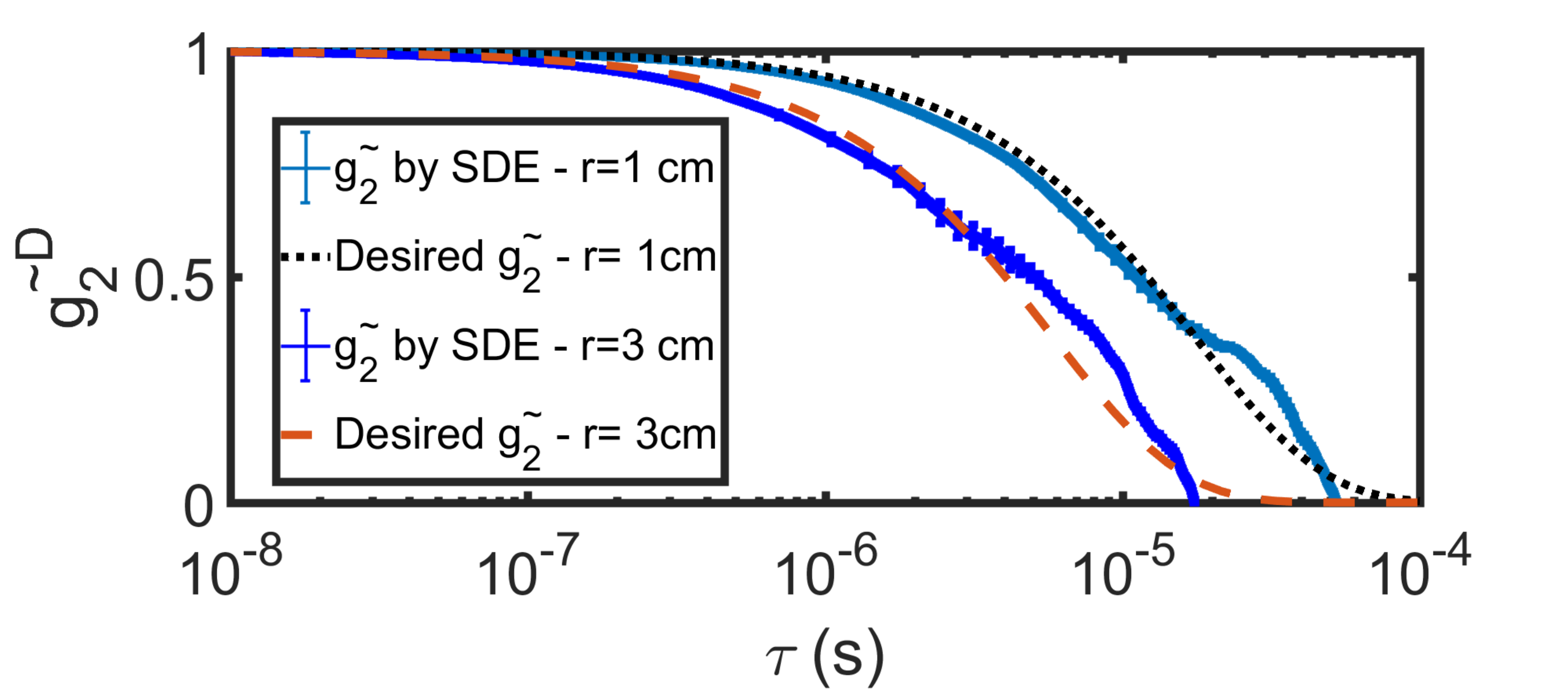}} 
\subfloat[]{\includegraphics[width = 3in]{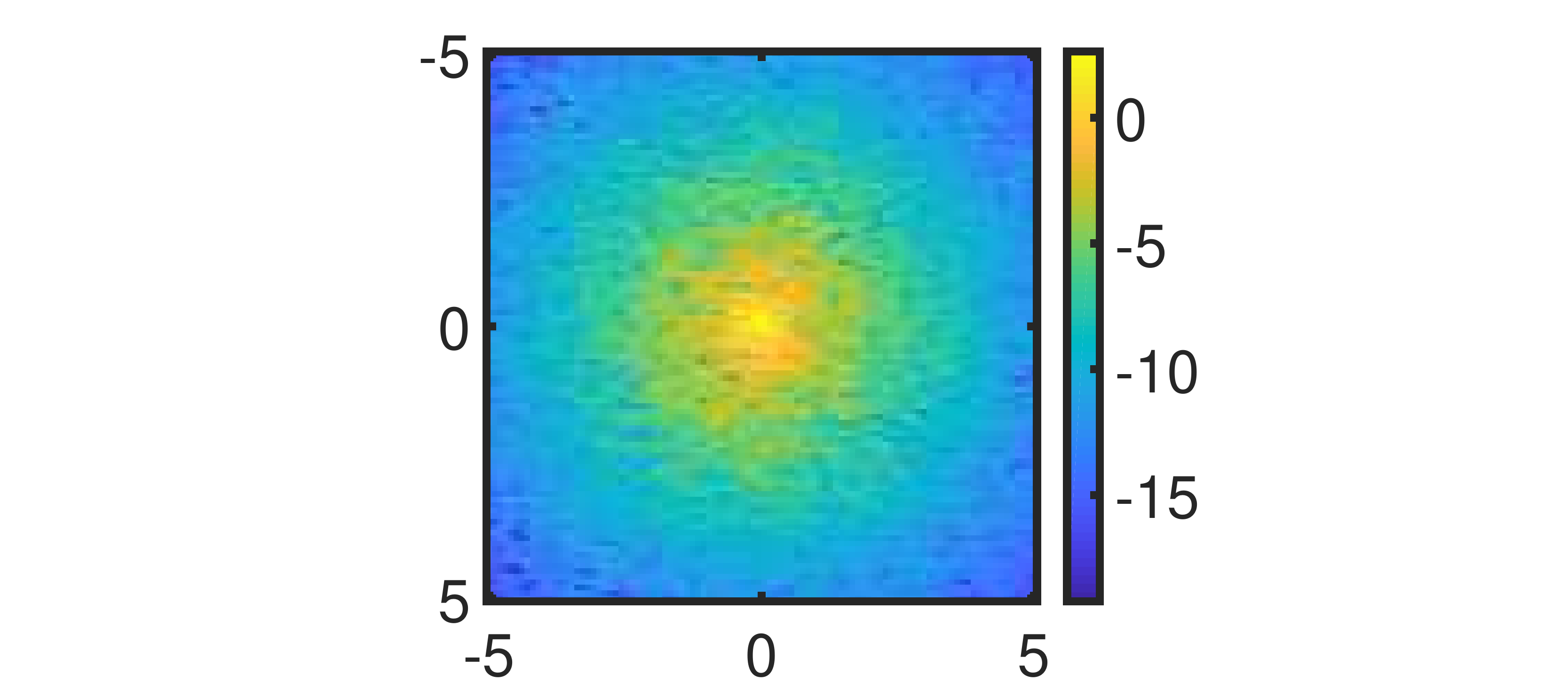}} \\
\subfloat[]{\includegraphics[width = 3in]{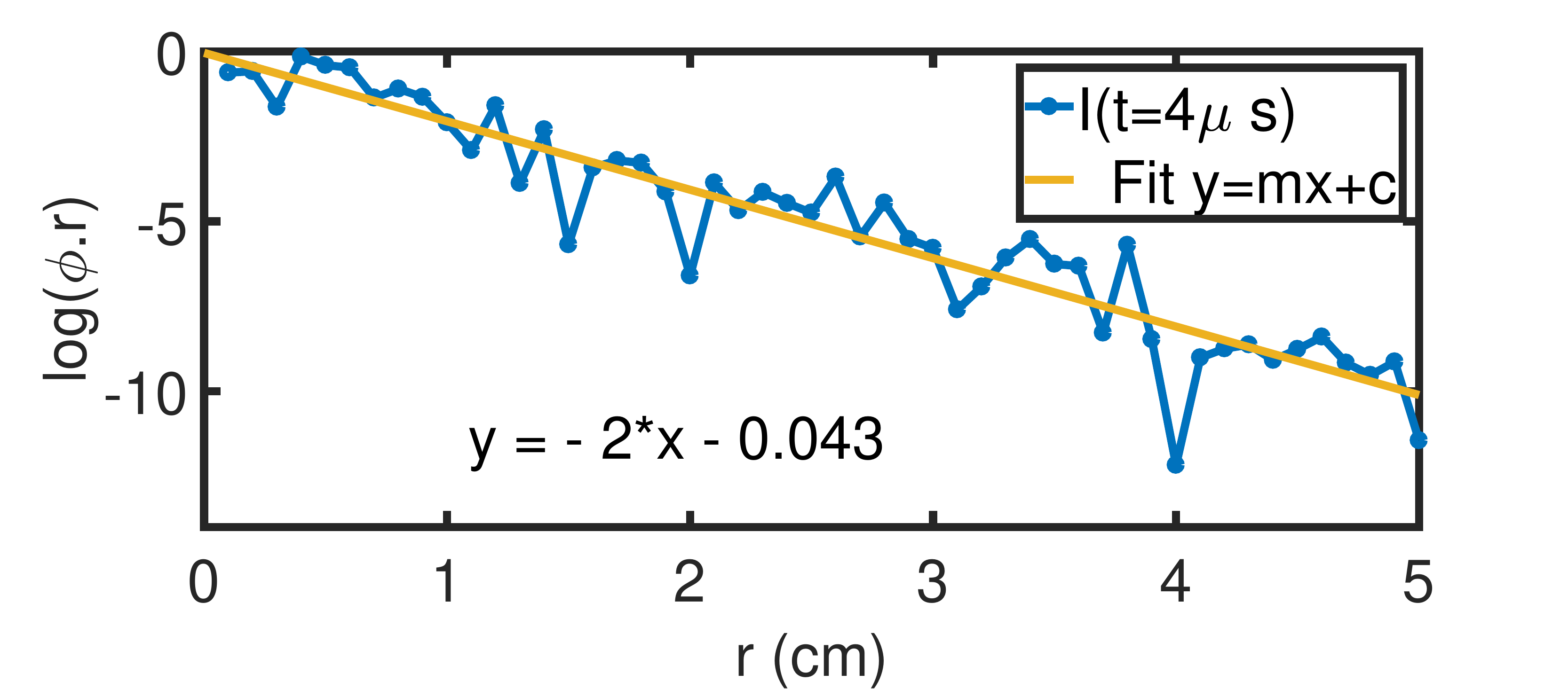}}
\subfloat[]{\includegraphics[width = 3in]{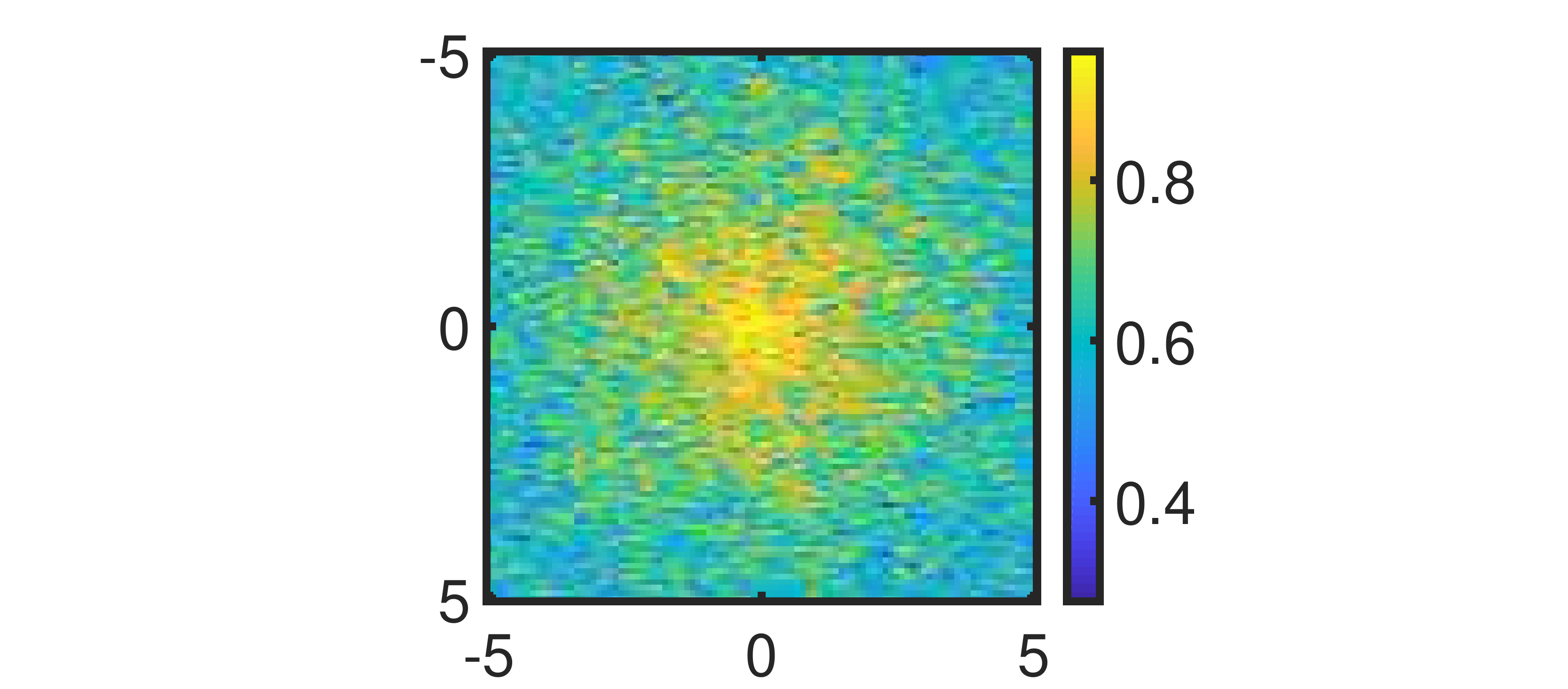}}\\
\subfloat[]{\includegraphics[width = 3in]{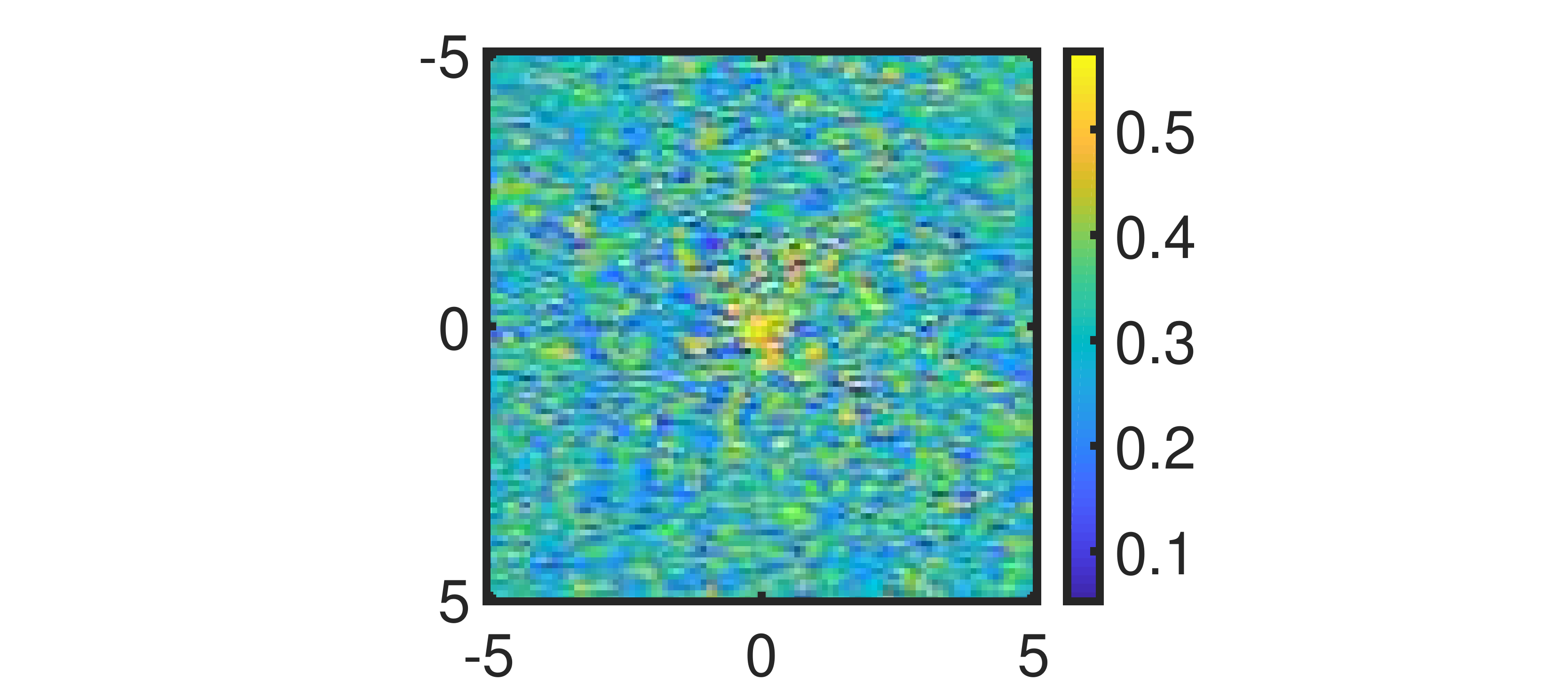}} 
\caption{Figure (a) shows the  $\tilde{g_2}^D$ obtained using SDE for two different r values. It can be seen that $\tilde{g_2}^D$ obtained using SDE shifts to left as r increases and is compared with the desired $\tilde{g_2}^D$. Figure (b) shows the log of intensity speckles obtained as an image with a point source at the centre for a multiple scattering model. The intensity profile (i.e.,$log(\phi.r)$) is plotted as function of r is shown in Fig (c) and its slope is shown. Figures (d) and (e) corresponds to auto-correlations $\tilde{g_2}^D$ obtained at $\tau=4\mu s$ and $\tau=40 \mu s$  respectively.}
\label{numerical2}
\end{figure}

One of the current limitations of the method is that as $t_{max}$ increases the auto-correlation curve deviates from the desired value. From the earlier analysis, the  $t_{max}$ should be around $3\times 1/\tilde{\alpha}$. For higher values of $t_{max}$, the auto-correlation curve obtained is shown in Fig 4(a). This limitation has to be fixed for a better estimate for $I(t)$ to be obtained at higher values of $t$. In order to increase $t_{max}=1000/\tilde{\alpha}$, we had initialized the initial solution $I_0$ as given in equation (5) to be $1000 \times \mu$. This resulted in better estimate of auto-correlation curve as shown in Fig 4(c), but the intensity I(t) has a transient component as seen from Fig 4(b). The corresponding histogram is shown in Fig 4(d). Although the histogram and auto-correlation curves are well-behaved, the intensity speckle has transient nature for the initial values of time t, which needs to be further addressed. Additionally, incorporating more complex autocorrelation, without need of binomial approximation, along with semi-infinite correlation diffusion solution has to be further explored.

\begin{figure}
\subfloat{\includegraphics[width = 3in]{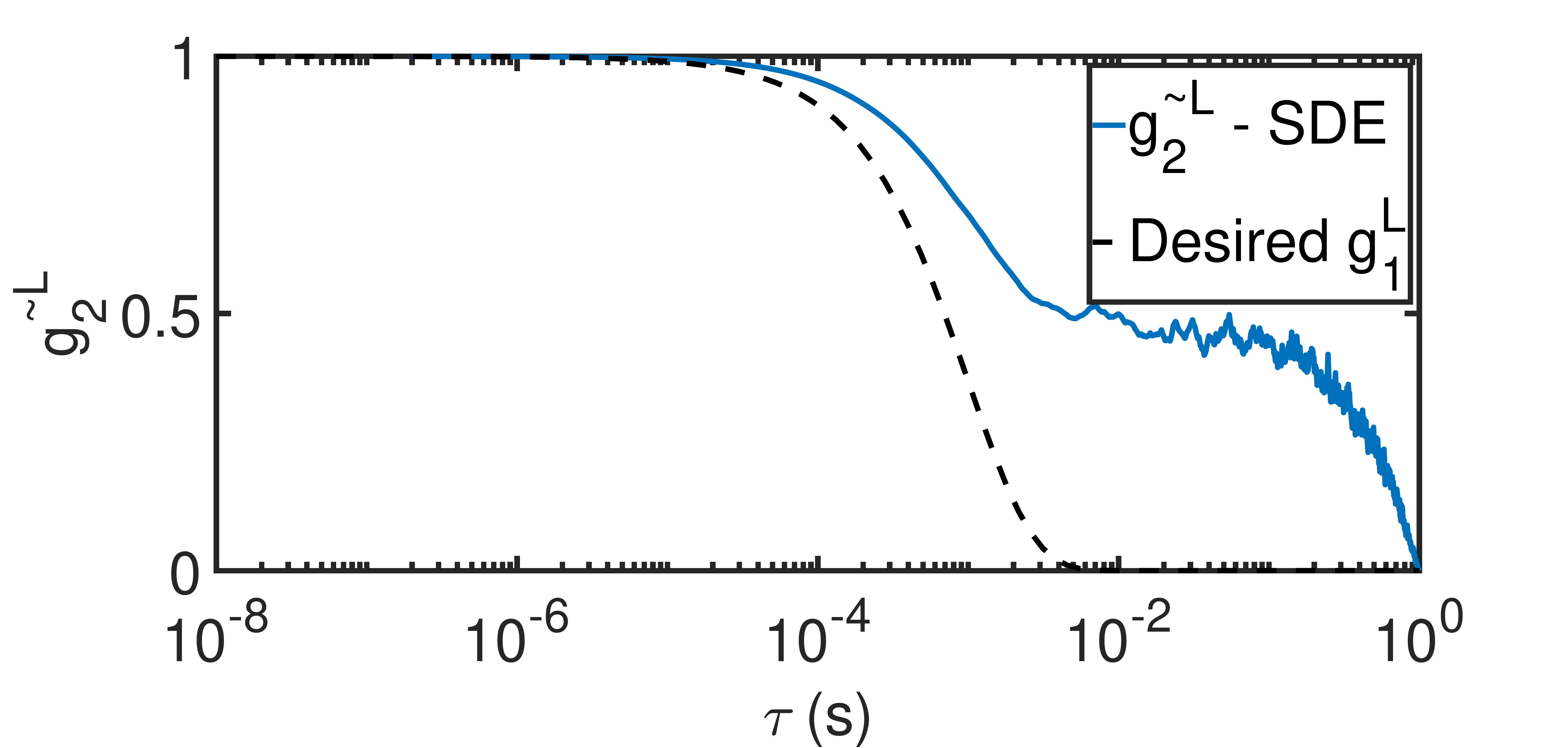}} 
\subfloat{\includegraphics[width = 3in]{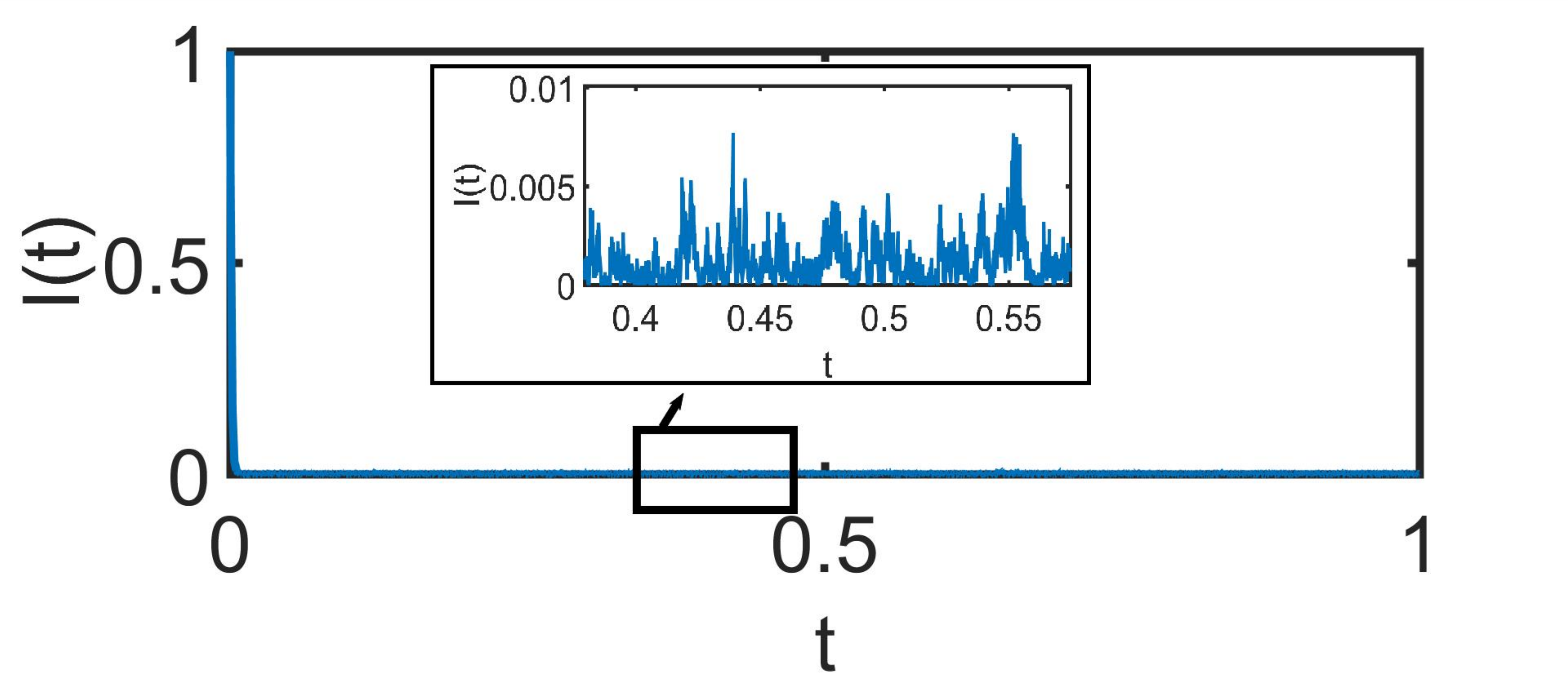}} \\
\subfloat{\includegraphics[width = 3in]{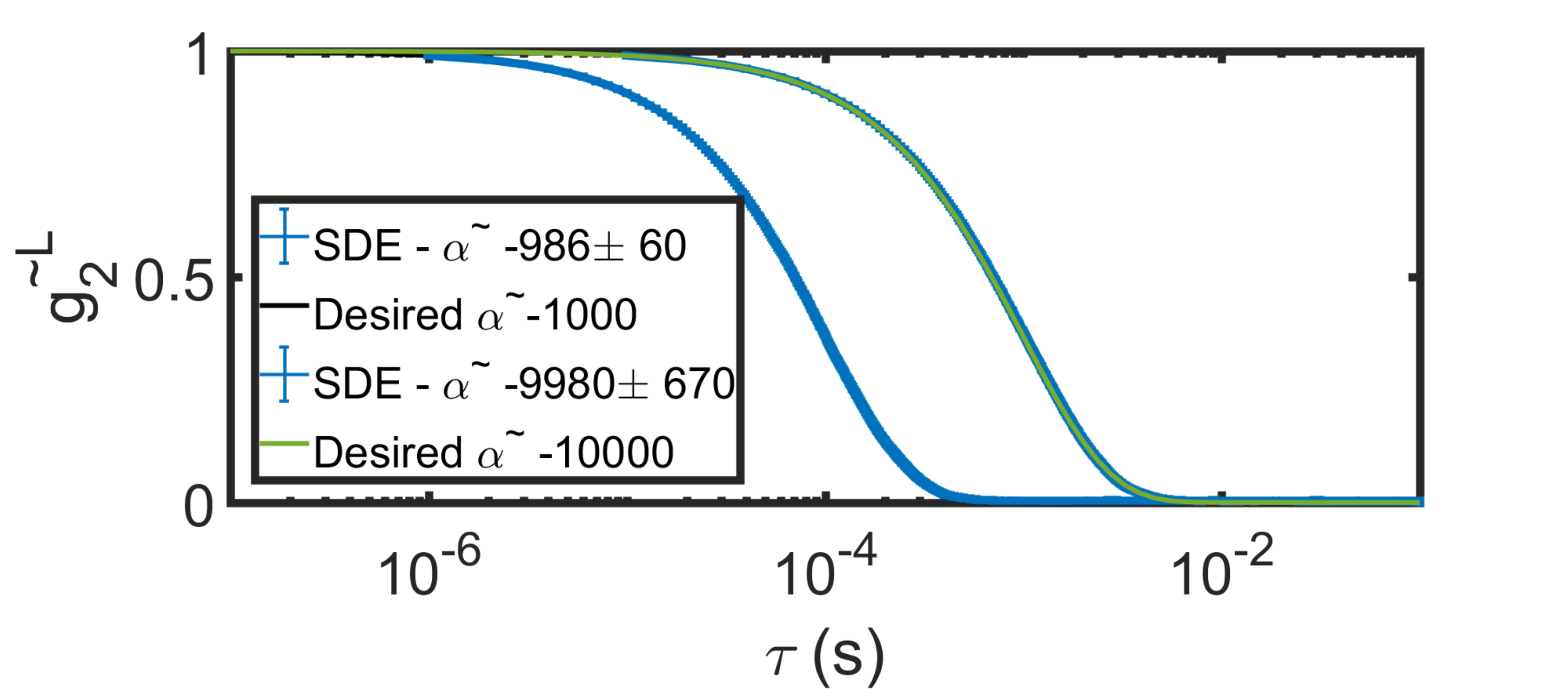}}
\subfloat{\includegraphics[width = 3in]{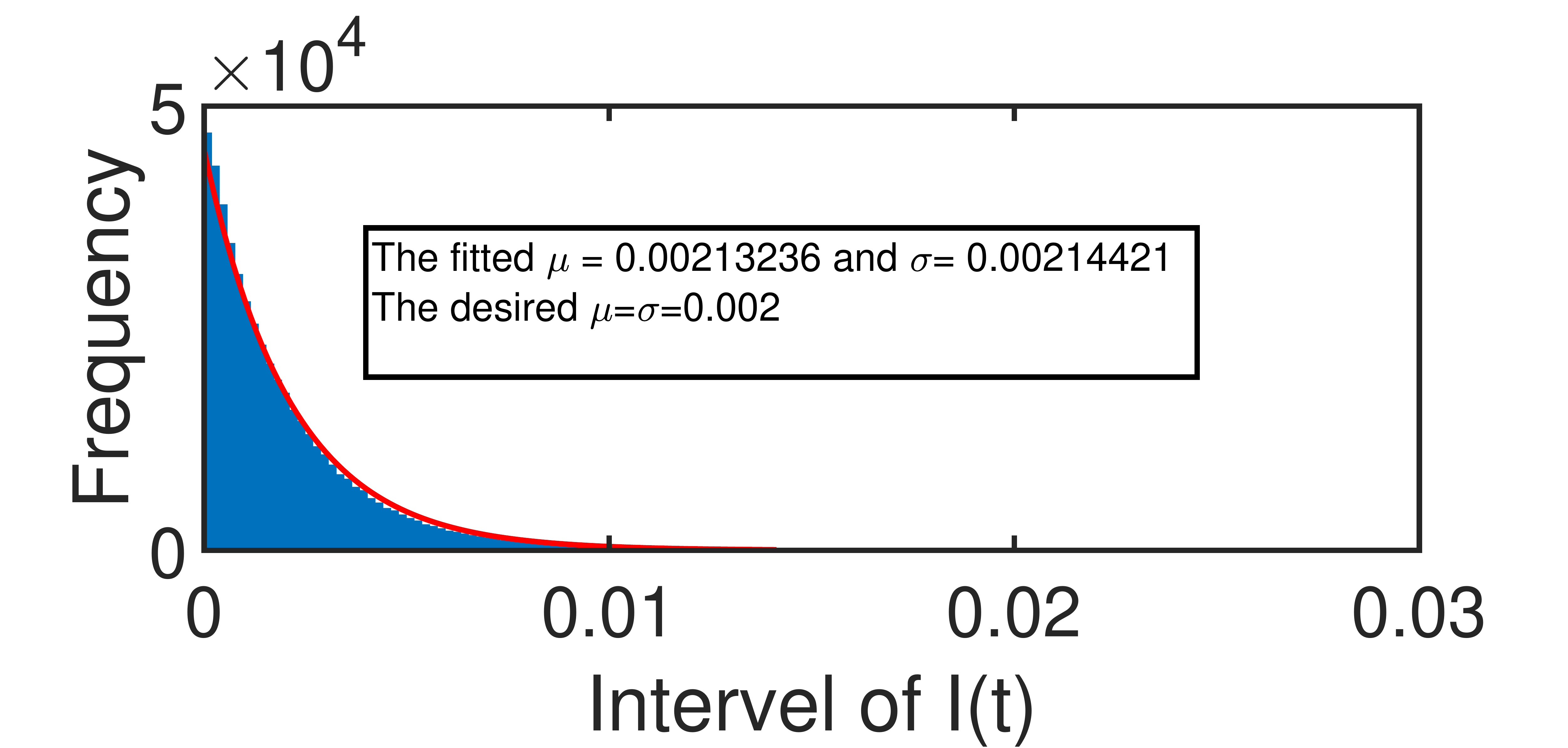}}\\
\caption{Figures (a) shows the $\tilde{g_2}^L$ obtained using SDE obtained when $t_{max}=1000\times1/\tilde{\alpha}$ and when $I_0=\mu$. Figure (b) shows the intensity speckles obtained when initial value was fixed as $1000\times \mu$ and when $t_{max}=1000 / \tilde{\alpha}$. It can seen that there is an initial transient and it settles after some time. The corresponding auto-correlation and histogram and shown in Figure (c) and (d) respectively. }
\label{numerical3}
\end{figure}

\printbibliography

@article{boas2010laser,
  title={Laser speckle contrast imaging in biomedical optics},
  author={Boas, David A and Dunn, Andrew K},
  journal={Journal of biomedical optics},
  volume={15},
  number={1},
  pages={011109},
  year={2010},
  publisher={International Society for Optics and Photonics}
}

@article{durduran2010diffuse,
  title={Diffuse optics for tissue monitoring and tomography},
  author={Durduran, Turgut and Choe, Regine and Baker, WB and Yodh, Arjun G},
  journal={Reports on Progress in Physics},
  volume={73},
  number={7},
  pages={076701},
  year={2010},
  publisher={IOP Publishing}
}

@article{briers1982retinal,
  title={Retinal blood-flow visualization by means of laser speckle photography.},
  author={Briers, JD and Fercher, AF},
  journal={Investigative ophthalmology \& visual science},
  volume={22},
  number={2},
  pages={255--259},
  year={1982},
  publisher={The Association for Research in Vision and Ophthalmology}
}

@article{parthasarathy2008robust,
  title={Robust flow measurement with multi-exposure speckle imaging},
  author={Parthasarathy, Ashwin B and Tom, W James and Gopal, Ashwini and Zhang, Xiaojing and Dunn, Andrew K},
  journal={Optics express},
  volume={16},
  number={3},
  pages={1975--1989},
  year={2008},
  publisher={Optical Society of America}
}

@article{bi2013deep,
  title={Deep tissue flowmetry based on diffuse speckle contrast analysis},
  author={Bi, Renzhe and Dong, Jing and Lee, Kijoon},
  journal={Optics letters},
  volume={38},
  number={9},
  pages={1401--1403},
  year={2013},
  publisher={Optical Society of America}
}

@article{valdes2014speckle,
  title={Speckle contrast optical spectroscopy, a non-invasive, diffuse optical method for measuring microvascular blood flow in tissue},
  author={Valdes, Claudia P and Varma, Hari M and Kristoffersen, Anna K and Dragojevic, Tanja and Culver, Joseph P and Durduran, Turgut},
  journal={Biomedical optics express},
  volume={5},
  number={8},
  pages={2769--2784},
  year={2014},
  publisher={Optical Society of America}
}

@article{murali2019recovery,
  title={Recovery of the diffuse correlation spectroscopy data-type from speckle contrast measurements: towards low-cost, deep-tissue blood flow measurements},
  author={Murali, K and Nandakumaran, AK and Durduran, Turgut and Varma, Hari M},
  journal={Biomedical optics express},
  volume={10},
  number={10},
  pages={5395--5413},
  year={2019},
  publisher={Optical Society of America}
}

@article{murali2020multi,
  title={Multi-speckle diffuse correlation spectroscopy to measure cerebral blood flow},
  author={Murali, K and Varma, Hari M},
  journal={Biomedical Optics Express},
  volume={11},
  number={11},
  pages={6699--6709},
  year={2020},
  publisher={Optical Society of America}
}

@article{zhou2018highly,
  title={Highly parallel, interferometric diffusing wave spectroscopy for monitoring cerebral blood flow dynamics},
  author={Zhou, Wenjun and Kholiqov, Oybek and Chong, Shau Poh and Srinivasan, Vivek J},
  journal={Optica},
  volume={5},
  number={5},
  pages={518--527},
  year={2018},
  publisher={Optical Society of America}
}

@article{kirkpatrick2008detrimental,
  title={Detrimental effects of speckle-pixel size matching in laser speckle contrast imaging},
  author={Kirkpatrick, Sean J and Duncan, Donald D and Wells-Gray, Elaine M},
  journal={Optics letters},
  volume={33},
  number={24},
  pages={2886--2888},
  year={2008},
  publisher={Optical Society of America}
}

@article{duncan2008copula,
  title={The copula: a tool for simulating speckle dynamics},
  author={Duncan, Donald D and Kirkpatrick, Sean J},
  journal={JOSA A},
  volume={25},
  number={1},
  pages={231--237},
  year={2008},
  publisher={Optical Society of America}
}

@article{song2016simulation,
  title={Simulation of speckle patterns with pre-defined correlation distributions},
  author={Song, Lipei and Zhou, Zhen and Wang, Xueyan and Zhao, Xing and Elson, Daniel S},
  journal={Biomedical optics express},
  volume={7},
  number={3},
  pages={798--809},
  year={2016},
  publisher={Optical Society of America}
}

@book{rabal2018dynamic,
  title={Dynamic laser speckle and applications},
  author={Rabal, Hector J and Braga Jr, Roberto A},
  year={2018},
  publisher={CRC press}
}

@article{yin2015improved,
  title={An improved Milstein method for stiff stochastic differential equations},
  author={Yin, Zhengwei and Gan, Siqing},
  journal={Advances in Difference Equations},
  volume={2015},
  number={1},
  pages={1--16},
  year={2015},
  publisher={Springer}
}

@book{goodman2007speckle,
  title={Speckle phenomena in optics: theory and applications},
  author={Goodman, Joseph W},
  year={2007},
  publisher={Roberts and Company Publishers}
}

@article{higham2001algorithmic,
  title={An algorithmic introduction to numerical simulation of stochastic differential equations},
  author={Higham, Desmond J},
  journal={SIAM review},
  volume={43},
  number={3},
  pages={525--546},
  year={2001},
  publisher={SIAM}
}

@book{braumann2019introduction,
  title={Introduction to stochastic differential equations with applications to modelling in biology and finance},
  author={Braumann, Carlos A},
  year={2019},
  publisher={John Wiley \& Sons}
}

@article{zarate2016construction,
  title={Construction of SDE-based wind speed models with exponentially decaying autocorrelation},
  author={Z{\'a}rate-Mi{\~n}ano, Rafael and Milano, Federico},
  journal={Renewable Energy},
  volume={94},
  pages={186--196},
  year={2016},
  publisher={Elsevier}
}

@misc{gardinerstochastic,
  title={Stochastic methods: a handbook for the natural and social sciences 4th ed.(2009)},
  author={Gardiner, C},
  publisher={Springer}
}

\end{document}